\documentclass[11pt,twoside]{article}
\usepackage[pdftex]{graphicx}  
\usepackage{amsmath}
\usepackage{amssymb}
\usepackage{cite}

\begin{document}
\newcommand{\pst}{\hspace*{1.5em}}

\newcommand{\rigmark}{\em Journal of Russian Laser Research}
\newcommand{\lemark}{\em Volume 30, Number 5, 2009}

\newcommand{\be}{\begin{equation}}
\newcommand{\ee}{\end{equation}}
\newcommand{\bm}{\boldmath}
\newcommand{\ds}{\displaystyle}
\newcommand{\bea}{\begin{eqnarray}}
\newcommand{\eea}{\end{eqnarray}}
\newcommand{\ba}{\begin{array}}
\newcommand{\ea}{\end{array}}
\newcommand{\arcsinh}{\mathop{\rm arcsinh}\nolimits}
\newcommand{\arctanh}{\mathop{\rm arctanh}\nolimits}
\newcommand{\bc}{\begin{center}}
\newcommand{\ec}{\end{center}}

\thispagestyle{plain}

\label{sh}


\begin{center} {\Large \bf
\begin{tabular}{c}
WAVE FUNCTION OF CLASSICAL PARTICLE
\\[-1mm]
IN LINEAR POTENTIAL
\end{tabular}
 } \end{center}

\bigskip

\bigskip

\begin{center} {\bf
A. S. Avanesov,$^{1*}$ V. I. Manko$^2$ }\end{center}

\medskip

\begin{center}
{\it
$^1$Department of General and Applied Physics, Moscow Institute of Physics and Technology\\
Moscow, Russia

\smallskip

$^2$P.N. Lebedev Physical Institute, Russian Academy of Sciences\\
Leninskii Prospect, 53, Moscow 119991, Russia }
\smallskip

$^*$Corresponding author e-mail:~~~daypatu~@~rambler.ru\\
\end{center}

\begin{abstract}\noindent
The problem of classical particle in linear potential is studied by
using the formalism of Hilbert space and tomographic probability
distribution. The Liouville equation for this problem is solved by
finding the density matrix satisfying von Newmann-like equation in
the form of  product of wave functions. The relation to quantum
mechanics is discussed.
\end{abstract}

\medskip


\section{Introduction}
\pst Recently \cite{1} the tomographic probability representation of
quantum states was introduced. In the probability representation the
states are identified with the fair probability distributions
connected with wave functions by integral transforms
\cite{2}\cite{3}. The density matrix of mixed quantum states can be
also related to the tomographic probability representation \cite{4}.

The Wigner functions \cite{5} is determined  by the tomographic
probability distribution in view of Radon transform \cite{6}. In
\cite{7} the tomographic probability distribution was introduced to
describe the states of classical particles. This approach was based
on using the Radon transform of the probability distribution
function $f(q,p,t)$ on the phase space. The tomographic probability
was called the state tomogram and its properties were discussed in
\cite{8} in the context of studying the classical and quantum chaos.
Thus, there exists the duality of considering classical and quantum
systems. One can formulate quantum mechanics by using the
mathematical tools of classical statistics like probability
distribution functions considered as alternative of wave functions.
On the other hand, one can introduce the quantum-like formalism of
Hilbert space, wave functions and density operators considering
classical systems. The idea of the introducing the Hilbert space
formalism for classical system states was suggested in \cite{9}. But
in the tomographic approach the quantum-like formalism used in
classical domain is different from suggested by Koopman, though the
general idea is related to the approach presented in \cite{9} and
other works \cite{10}\cite{11}. Some relations of quantum-like and
classical descriptions were studied in \cite{17}\cite{22}.

 Tomographic representation of quantum and classical mechanics was reviewed in \cite{12}.
It was shown that quantum states with discrete variables like spin
also can be described by using probability distribution functions
called tomograms \cite{13}\cite{14}.  For continuous variables we
got an idea of describing classical states by using formalism of
quantum mechanics, but for discrete variables such approach is not
developed yet. Thus, in our work we concentrate on introducing
description of the states of classical systems with continuous
variables by using formalism of quantum mechanics and following
\cite{15}.

In this work we consider one example of classical systems. It is the
problem of a particle moving in the linear potential. On this
example we consider some connections between the standard
description and Hilbert space representations of classical
mechanics.

The paper is organized as follows. In next section Sec. 2 we
describe our classical system using standard approach. Then,in Sec.
3 we introduce density matrix by using the Wigner function and its
similarity to classical probability distribution function
$f(q,p,t)$. In next two sections, Sec. 4 and Sec. 5, classical wave
function for gaussian state is found and also relation between
evolution operator of Liouville equation and Green's function of
wave equation is discussed. The conclusions are presented in Sec. 6.

\section{Review of the classical problem}
\pst Let us formulate our problem. We consider classical particle
motion in the field with linear potential $V=q$. Here $q$ is a
position of the classical particle. We use positive sign for
potential. For instance, our problem can be considered as the
problem of electron's moving between plates of flat condenser. Also
we take the mass of our particle and its charge $m=Q=1$. Thus, we
take the Hamiltonian of particle moving in linear potential in the
form \be
  H=\frac{p^2}{2}+q .
\ee So, we get the equations of motions for the position q and the
momentum p
$$ \frac{\partial H}{\partial p}=\dot{q}, -\frac{\partial H}{\partial q}=\dot{p}  .$$
That yields
 \be
  \ddot{q}+1=0  .
\ee It is easy to solve this equation, so let us write the solution

\be
 \begin{gathered}
  p=p_{0}-t  ,\\
  q=q_{0}+p_{0}t-\frac{t^2}{2}  .
 \end{gathered}
\ee
Here $p_{0}$ is initial momentum, $q_{0}$ is initial position.

In case position and momentum fluctuate, the state of our system is
described by a probability distribution function $f(q,p,t)$ on the
phase space. This function is nonnegative and it satisfies the
normalization condition, i. e.

\be
 \begin{gathered}
  f\left(q,p,t\right)\geq0  ,\\
  \int f(q,p,t) dq dp=1  .
 \end{gathered}
\ee

The function is a solution of the Liouville kinetic equation
$$\frac{\partial f}{\partial t}+\frac{\partial H}{\partial p}\frac{\partial f}{\partial q}-\frac{\partial H}{\partial q}\frac{\partial f}{\partial p}=0  .$$
In case of linear potential the Liouville equation reads
 \be
  \frac{\partial f}{\partial
t}+p\frac{\partial f}{\partial q}-\frac{\partial f}{\partial p}=0  .
\ee
In next sections we will consider solutions of the above
equation by applying the quantum-like formalism of the Hilbert space
vectors and the corresponding wave functions.

\section{Classical wave function}
\pst

So, we want to consider solution of equation (5) by using
quantum-like formalism. But before we start we should show how we
introduce this formalism in classical mechanics.
 In quantum mechanics,
the states are described by a wave function or a density matrix. But
also we can use other representations. For instance there exists the
phase-space quasidistribution $W(q,p,t)$ called the Wigner function.
It is given by the Fourier transform of a density matrix
$\rho(x,x',t)$ introduced in \cite{16}
 \be
  W(q,p,t)=\int\rho\left(q+\frac{u}{2},q-\frac{u}{2},t\right)e^{-ipu}du
  .
\ee The inverse Fourier transform yields
 \be
  \rho(x,x',t)=\frac{1}{2\pi}\int
W\left(\frac{x+x'}{2},p,t\right)e^{ip(x-x')}dp  .\ee

So we get the relation between this two representations of quantum
mechanics. They provide two ways to describe quantum states. Doing
the transform (6) we associate the quantum state with the function
on phase space.

In classical mechanics we are usually (we can say always) work in
phase space. It is the standard tool. But it is possible to make
transform to use the Hilbert space representation by introducing
density matrix (density operator) given formally by (7).
 So, let us consider probability distribution $f(q,p,t)$ which is an analog of the Wigner function.
 It means that we replace $W(q,p,t)$ by $2\pi f(q,p,t)$ in formula (7). Then we can
write expression for density matrix of classical motion
 \be
\rho(x,x',t)=\int f\left(\frac{x+x'}{2},p,t\right)e^{ip(x-x')} dp
 .\ee

Then it is important to introduce the evolution equation. In our
case we get the equation which resembles the quantum equation. In
fact, doing some transformations with Liouville equation (5),
finally we get von Neumann-like equation for the density matrix
 \be
  i\frac{\partial
\rho\left(x,x',t\right)}{\partial
t}=-\frac{1}{2}\left(\frac{\partial^2 }{\partial
x^2}-\frac{\partial^2 }{\partial
x'^2}\right)\rho\left(x,x',t\right)+(x-x')\rho\left(x,x',t\right)  ,
\ee
where we used the substitution rules
\be
 \begin{gathered}
  \frac{\partial f}{\partial t}\rightarrow\frac{\partial
\rho\left(x,x',t\right)}{\partial t}   ,\\
  \frac{\partial f}{\partial p}\rightarrow\-i(x-x')\rho(x,x't)  ,\\
  p\frac{\partial f}{\partial
q}\rightarrow\frac{1}{2i}\left(\frac{\partial^2 }{\partial
x^2}-\frac{\partial^2 }{\partial x'^2}\right)\rho(x,x',t)  .
 \end{gathered}
\ee

This equation is equation with separable variables $x$, $x'$. In
view of this one can find the solution of the equation factorizing
the density matrix in the form
 \be
  \rho\left(x,x',t\right)=\Psi\left(x,t\right)\Psi^*\left(x',t\right)
  .
\ee
Then, one can show that the function $\Psi(x,t)$, which can be
called the wave function of classical particle satisfies the
$\textrm{Shr\"{o}dinger-like}$ equation.

But let us pay our attention on the following circumstance. The
formula for the density matrix provides the possibility to make the
gauge transformation of the wave function
 \be
  \Psi(x,t)\rightarrow\Psi(x,t)e^{i\phi(t)}  .
\ee Here $\phi(t)$ is a real phase factor depending on time. From
this point of view, the classical probability density $f(q,p,t)$ can
be mapped onto density matrix of the factorized form
$\rho(x,x',t)=\Psi(x,t)\Psi^*(x',t)$. The corresponding wave
function is determined up of this gauge factor. But we want the
function $\Psi(x,t)$ to satisfy $\textrm{Shr\"{o}dinger-like}$
equation. So, when we use this condition we get equation for the
wave function $\Psi(x,t)$

\be
  i\frac{\partial \Psi\left(x,t\right)}{\partial
t}=-\frac{1}{2}\frac{\partial^2 \Psi\left(x,t\right)}{\partial
x^2}+x\Psi\left(x,t\right)  .
\ee

It is worthy to point on that we lose time-dependent phase-factor
when we get $\textrm{Shr\"{o}dinger-like}$ equation from  the von
Neumann-like equation for density matrix.

Also we must take into account that not each solution of wave
equation provides the real probability distribution function. The
function $f(q,p,t)$ expressed in terms of solution $\Psi(x,t)$ of
$\textrm{Shr\"{o}dinger-like}$ always satisfies Liouville equation,
but this solution can take negative values which is not admissible
for classical probability distribution.


\section{Gaussian solution}
\pst

In this section we will consider gaussian solution of our problem
and show some connections between these two representations. Firstly
we will find solution of the Liouville equation, in another words we
will get probability density function and using (7) we will find the
density matrix.
 The Liouville equation can be reduced, by using the Fourier
transform, to the von Neumann-like equation as it was demonstrated.
Let us suppose that the probability distribution at $t=0$ has the
Gaussian form
 \be
  f_{0}\left(q_{0},p_{0}\right)=f\left(q,p,0\right)=\frac{1}{\pi}e^{-q_{0}^2-p_{0}^2}
  .
\ee

We want to find evolution of the probability distribution function.
This problem can be solved easily.

In fact, the equations of motion read
$$p=p_{0}-t  ,$$
$$q=q_{0}+p_{0}t-\frac{t^2}{2}  .$$

For the initial momentum and the initial position we find the
expressions
 \be
 \begin{gathered}
  p_{0}=p+t  ,\\
  q_{0}=q-pt-\frac{t^2}{2}  .
 \end{gathered}
\ee Since the probability density is the integral of motion from the
Liouville theorem we get the probability distribution function for
arbitrary time moment
 \be
  f\left(q,p,t\right)=\frac{1}{\pi}e^{-\left(p+t\right)^2-\left(q-pt-\frac{t^2}{2}\right)^2}
  .
\ee

Also we can find the evolution of statistical parameters of our
Gaussian probability distribution like variances and covariance of
random position and momentum.

We have the initial dispersion matrix
 \be
  C_{0}=\frac{1}{2}
     \begin{Vmatrix} 1&0\\0&1\end{Vmatrix}  .
\ee

The transformation matrix, corresponding to the evolution (16) reads

\be
  A=\begin{Vmatrix} 1&t\\0&1\end{Vmatrix}  .
\ee

Then we get the dispersion matrix for arbitrary time moment $t$
 \be
  C=AC_{0}A^T=\frac{1}{2}\begin{Vmatrix} 1+t^2&t\\t&1\end{Vmatrix}
  .
\ee

Our aim is to find the wave function that is associated with this
probability distribution function. We put expression for $f(q,p,t)$
(16) in (8), and get the density matrix
 \be
  \rho\left(x,x',t\right)=\frac{1}{\pi}\int
e^{-\left(p+t\right)^2-\left(q-pt-\frac{t^2}{2}\right)^2}e^{ip\left(x-x'\right)}
dp  .
\ee

Factorizing the expression for the density matrix $\rho(x,x',t)$ we
get this wave function $\Psi(x,t)$
 \be
  \Psi\left(x,t\right)=\frac{\exp\left\{-\frac{1}{4}\frac{t^2/2+2x^2+2t^2x+4ixt-2ix^2t+2it^3x}{(1+t^2)}\right\}}{\sqrt[4]{\pi\left(1+t^2\right)}}
  .
\ee

There is the problem of the time-dependent phase-factor as pointed
out in the end of Sec. 3. That's why we have one problem, this wave
function (21) doesn't satisfy $\textrm{Shr\"{o}dinger-like}$
equation. So, we can see it
 \be
  \left(-\frac{1}{2}\frac{\partial^2 }{\partial
x^2}-i\frac{\partial }{\partial
t}+x\right)\Psi\left(x,t\right)=\frac{1}{8}\left(
\frac{4+8t^2+3t^4+t^6}{\left(1+t^2\right)^2}\right)\Psi\left(x,t\right)\neq0
.
 \ee

But we want the wave function of classical particle to satisfy the
$\textrm{Shr\"{o}dinger-like}$ equation. Because of the gauge
invariance, in general, the wave function has the form
 \be
  \Psi\left(x,t\right)=\frac{\exp\left\{-\frac{1}{4}\frac{t^2/2+2x^2+2t^2x+4ixt-2ix^2t+2it^3x}{1+t^2}+i\phi\left(t\right)\right\}}{\sqrt[4]{\pi\left(1+t^2\right)}}
.
 \ee

As we require, this function has to satisfy
$\textrm{Shr\"{o}dinger-like}$ equation (13). Using this condition
and putting (23) in (13), we find the equation for time dependent
phase $\phi(t)$
 \be
  \left(
\frac{4+8t^2+3t^4+t^6}{8\left(1+t^2\right)^2}+\dot{\phi\left(t\right)}\right)\Psi\left(x,t\right)=0
.
 \ee

Solving the differential equation for the phase $\phi(t)$, we
finally get
 \be
  \phi\left(t\right)=-\frac{t}{8}-\frac{t^3}{24}-\frac{1}{2}\arctan\left(t\right)+\frac{1}{8}\left(\frac{t}{1+t^2}\right)+const
.
\ee

In our case, for this choice of time-dependent phase, the wave
function $\Psi(x,t)$ satisfies the $\textrm{Shr\"{o}dinger-like}$
equation (13).

\section{Relation between propagator of the Liouville equation and Green function of the wave equation}
\pst

In previous section we found gaussian solution of our classical
problem. We presented two ways to describe our classical system. We
found probability distribution (16) and corresponding wave function
(23). Also we can find this wave function using Green function of
the $\textrm{Shr\"{o}dinger-like}$ equation. In this section we want
get it.

 We are going to find the relation between the propagator of the Liouville
equation (5) and the Green function of the
$\textrm{Shr\"{o}dinger-like}$ equation (13).

 Let us introduce the following maps:
\be
  \rho\stackrel{\hat{F}}{\rightarrow}f,\;
  \rho_{0}\stackrel{\hat{F}}{\rightarrow}f_{0},\;
  f_{0}\stackrel{\hat{\Pi}}{\rightarrow}f,\;
  \rho_{0}\stackrel{\hat{B}}{\rightarrow}\rho  .
\ee
The operators $\hat{F}$, $\hat{F^{-1}}$, $\hat{\Pi}$, $\hat{B}$
can be given in matrix form expressed as kernels of the integral
transforms. These maps are determined by the following expressions

$$ \rho(x,y,t)=\int F^{-1}(x,y,q,p,t) f(q,p,t) dq dp $$   and
$$ f(q,p,t)=\int F(x,y,q,p,t) \rho(x,y,t) dx dy  ,$$
here from (6), (7), (8) we find
$$F^{-1}(x,y,q,p,t)=e^{ip(x-y)}\delta\left(\frac{x+y}{2}-q\right),$$
$$F(x,y,q,p,t)=\frac{1}{2\pi} e^{-ip(x-y)}\delta\left(\frac{x+y}{2}-q\right)  ,$$

$$ f(q,p,t)=\int \Pi(q,p,q',p',t) f_{0}(q',p') dq' dp'   ,$$
here $\Pi(q,p,q',p',t)$ is propagator for the Liouville equation,

$$ \rho(x,y,t)=\int B(x,y,x',y',t) \rho_{0}(x',y') dx' dy'  ,$$
the matrix $\hat{B}$ is expressed in terms of the
$\textrm{Shr\"{o}dinger-like}$ equation
 $B\left(x,y,x',y',t\right)=G\left(x,x',t\right)G^*\left(y,y',t\right)$

Then we get identity
 \be
  \hat{B}\rho_{0}=\hat{F}^{-1}\hat{\Pi}\hat{F}\rho_{0} ,
\ee that provide this expression
 \be
\begin{aligned}
    \int B\left(x,y,x',y',t\right)\rho_{0}\left(x',y'\right) dx'   dy'=\\=\frac{1}{2\pi}\int F^{-1}(x,y,q,p,t) dq dp \int dq' dp' \Pi\left(q,p,q',p',t\right) \int F(x',y',q',p',t)\rho_{0}\left(x',y'\right) dx' dy' .
\end{aligned}
\ee


We compare both part of this identity and get expression for
function $B(x,y,x',y',t)$. So, after some transformations we find
expression for function $B(x,y,x',y',t)$

\be
  B\left(x,y,x',y',t\right)=\frac{1}{2\pi} \int
e^{ip(x-y)}e^{-ip'(x'-y')}\Pi\left(q,p,q',p',t\right)
\delta\left(q-\frac{x+y}{2}\right)
\delta\left(q'-\frac{x'+y'}{2}\right) dq dq' dp' dp .
\ee

Propagator for the Liouville equation (5) reads
\be
\Pi\left(q,p,q',p',t\right)=\delta\left(q'-q_{0}(q,p,t)\right)\delta\left(p'-p_{0}(q,p,t)\right)
  .
\ee

Then, we get
\be
  B(x,y,x',y',t)=\int \frac{1}{2\pi t}\exp
\left\{i\frac{\left(q-q'-\frac{t^2}{2}\right)}{t}(x-y)\right\}
\delta\left(q-\frac{x+y}{2}\right)
\delta\left(q'-\frac{x'+y'}{2}\right) dq dq'
 . \ee

That provides Green function
 \be
  G(x,x',t)=\frac{1}{\sqrt{2\pi
t}}\,e^{i\frac{\left(x-x'\right)^2-t^2\left(x+x'\right)}{2t}}
.\ee

In fact, because of gauge invariance Green function has the form \be
  G(x,x',t)=\frac{1}{\sqrt{2\pi
t}}\,e^{i\frac{\left(x-x'\right)^2-t^2\left(x+x'\right)}{2t}+i\phi(t)}
.\ee

Indeed, we remember that we lose the time-dependent phase. We
remember our demand that Green function like wave function must
satisfy the $\textrm{Shr\"{o}dinger-like}$ equation. So, using last
condition we can find the time-dependent phase up to a constant.
Expression for the time-dependent phase reads
 \be
  \phi(t)=-\frac{t^3}{24}+const
.\ee

Thus we find Green function of the $\textrm{Shr\"{o}dinger-like}$
equation. It gives us opportunity to get all possible wave functions
and these functions can describe or determine the classical states.
But we should take into account the nonnegativity property of
probability distribution on phase space. Not each wave function
provides us with a real probability distribution. Using transform
(6) we can find solution of the Liouville equation. But possibly
this solution will not satisfy conditions (4) (for instance this
function can take negative values) and that is why it will not be
probability distribution function, in another words it will not
describe real classical state. In general, solution of the
$\textrm{Shr\"{o}dinger}$ equation always provides us with solution
of the Liouville equation but it does not have to correspond to real
classical state.

\section{Conclusions}
\pst To resume we formulate the main results of our work. We have
constructed the solution of Liouville equation of a classical
particle moving in linear potential in the form of Fourier transform
of the product of two "wave functions". The wave functions satisfy
the  $\textrm{Shr\"{o}dinger-like}$ equations identical to
$\textrm{Shr\"{o}dinger}$ equations for quantum particle moving in
the linear potential. We demonstrated on this example that in
classical mechanics one can use formalism of Hilbert spaces and
formalism of quantum density operator. It is another example in
addition to examples considered previously in \cite{9} and \cite{15}
which provides the possibility to use the same formalism both in
classical and quantum mechanics.

\section*{Acknowledgments}
\pst The author acknowledges the financial support provided within
the Project RFBR~11-02-00456.

\end{document}